\documentclass[epj]{svjour}
\pdfoutput=1
\usepackage{float}
\usepackage{subfigure}
\usepackage{braket}
\usepackage{color}
\usepackage{slashed}
\usepackage{amsmath}
\usepackage{amssymb}
\usepackage{graphicx}
\usepackage{url}
\begin{document}
\title{Detecting {F}luorescent {D}ark {M}atter with X-ray lasers}
\author{Francesca Day\inst{1} \and Malcolm Fairbairn\inst{2}}
\institute{DAMTP, CMS, Wilberforce Road, University of Cambridge, Cambridge, CB3 0WA, UK, \email{francesca.day@maths.cam.ac.uk} \and Physics, King's College London, Strand, London WC2R 2LS, UK, \email{malcolm.fairbairn@kcl.ac.uk}}
\date{Received: date / Revised version: date}
\abstract{Fluorescent Dark Matter has been suggested as a possible explanation of both the 3.5 keV excess in the diffuse emission of the Perseus Cluster and of the deficit at the same energy in the central active galaxy within that cluster, NGC 1275.  In this work we point out that such a dark matter candidate can be searched for at the new X-ray laser facilities that are currently being built and starting to operate around the world.  We present one possible experimental set up where the laser is passed through a narrow cylinder lined with lead shielding.  Fluorescent Dark Matter would be excited upon interaction with the laser photons and travel across the lead shielding to decay outside the cylinder, in a region which has been instrumented with X-ray detectors. For an instrumented length of 7cm at the LCLS-II laser we expect $\mathcal{O} (1 - 10)$ such events per week for parameters which explain the astronomical observations.
}

\maketitle

\newcommand{\be}{\begin{equation}}
\newcommand{\ee}{\end{equation}}
\newcommand{\bea}{\begin{eqnarray}}
\newcommand{\eea}{\end{eqnarray}}
\newcommand{\mbb}{\mathbb}
\newcommand{\ti}{\times}
\newcommand{\half}{\frac{1}{2}}
\newcommand{\mc}{\mathcal}
\newcommand{\gsim}{\gtrsim}
\newcommand{\lsim}{\lesssim}
\newcommand{\vphi}{\varphi}
\newcommand{\fn}{\footnote}
\newcommand{\water}{H$_2$O}
\newcommand{\by}{\bar{y}}
\newcommand{\bz}{\bar{z}}
\newcommand{\bo}{\bar{1}}
\newcommand{\bt}{\bar{2}}
\newcommand{\te}{\tilde{e}}
\newcommand{\tga}{\tilde{\Gamma}}
\newcommand{\zo}{z_{1}}
\newcommand{\zt}{z_{2}}
\newcommand{\bzo}{\bar{z}_{1}}
\newcommand{\bzt}{\bar{z}_{2}}
\newcommand{\tile}{\tilde{e}}
\newcommand{\del}{\partial}
\newcommand{\tK}{{ K}}
\newcommand{\tgamma}{{\tilde{\gamma}}}
\newcommand{\ph}{\phantom}
\newcommand{\nn}{\nonumber}

\newcommand{\aj}{AJ}
\newcommand{\apj}{ApJ}
\newcommand{\apjl}{ApJ}
\newcommand{\apjlett}{ApJ}
\newcommand{\apjs}{ApJS}
\newcommand{\apjsupp}{ApJS}
\newcommand{\aap}{A\&A}
\newcommand{\astap}{A\&A}
\newcommand{\araa}{ARA\&A}
\newcommand{\aapr}{A\&A Rev.}
\newcommand{\aaps}{A\&AS}
\newcommand{\mnras}{MNRAS}



\section{Introduction}
The cosmological and astrophysical evidence for dark matter is very compelling \cite{Zwicky:1933gu,Zwicky:1937zza,Rubin:1970zza,Rubin:1980zd,Ostriker:1974lna,Sofue:2000jx,Clowe:2006eq,Ade:2015xua}, however as yet all attempts to identify the elusive particle itself have failed.  Such a lack of experimental evidence in the lab does not on its own make the cold dark matter hypothesis less likely.
However, there is no particular reason why dark matter should have strong gauge couplings to Standard Model particles. One appropriate response to the current situation is therefore to search on all frontiers possible for particle dark matter - wherever we can constrain the parameter space we should do so \cite{wimpreview,axionreview,sterilereview}.  Particular models where either relic abundance is explained more naturally or, as in this paper, the candidate might explain some astrophysical anomalies are particularly well motivated places to start looking. 

One intriguing possible signal of dark matter is the 3.5 keV X-ray line, which has been detected in several astronomical objects including the Perseus cluster \cite{Bulbul:2014sua,Boyarsky:2014jta}.  It is possible that this line is a misidentified line in the spectra of normal gases in the Standard Model \cite{bananas,boring}. The origin of this line could also be the decay of dark matter, in which case the mass of the dark matter particle is typically double the energy of the emitted photon.  Another interpretation of this line is that it might correspond to the energy gap between a ground and excited state of dark matter \cite{Finkbeiner:2014sja,14107766,160304859,160801684}.  In that situation, one would expect emission and/or absorption at that particular frequency, depending upon the situation.

The original observations of the 3.5 keV line in Perseus made with the {\it XMM-Newton} and {\it Chandra} X-ray telescopes are of the diffuse halo. In these studies, point sources such as the central active galaxy are removed as a matter of course. 
Follow up observations made by the {\it Hitomi} Satellite before its untimely demise find no evidence for an emission line at 3.5 keV \cite{Hitomi}. In fact, these observations show a small \emph{dip} at 3.5 keV at a local significance of approximately $2.5 \sigma$. Furthermore, a dip in the photon spectrum at 3.5 keV has been observed in the spectrum of the active galaxy NGC 1275 at the centre of the Perseus Cluster \cite{Berg:2016ese,160801684}. It has been pointed out that the small dip observed by {\it Hitomi} could be a result of partial cancellation of an excess in the diffuse cluster and a dip in the central active galaxy NGC 1275 \cite{160801684}. Unlike {\it XMM-Newton} and {\it Chandra}, {\it Hitomi} does not have sufficient angular resolution to resolve NGC 1275 from the diffuse cluster background.

Such behaviour, with both emission and absorption lines observed, would be expected from a model of Fluorescent Dark Matter (FDM) which  has recently been suggested to explain these anomalies at 3.5 keV \cite{160801684}. In FDM, the absorption feature at 3.5 keV results from resonant excitation of dark matter. The 3.5 keV line observed in the diffuse emission from galaxy clusters in such a situation arises from fluorescent re-emission. Similar models have been considered previously, for example in \cite{Profumo:2006im,Finkbeiner:2014sja,14031750,160304859}. In particular \cite{160304859} uses a more general version of the model given in equation \eqref{FDMlagrangian} to explain the 3.5 keV line, as we discuss further in section 2. The discovery of anomalous absorption and emission features at the same energy brings additional observational motivation to FDM models. 

Here we describe a ground-based experiment to test this hypothesis by searching for Fluorescent Dark Matter using 3.5 keV photons from an X-ray laser.  As the laser light passes through either air or vacuum, it is also passing through the local distribution of dark matter, essentially a collisionless gas whose  density and velocity dispersion is thought to be relatively well known in the vicinity of the Solar System \cite{Fairbairn:2012zs}.  If dark matter does fluoresce within the energy range scanned by the laser, then absorption and subsequent re-emission of photons will look like anomalous apparent scattering, above and beyond any other sources of scattering. As we show below, next generation X-Ray lasers such as the Linac Coherent Light Source II (LCLS-II) \cite{LCLSII} could provide an observable signal in such an experiment.

In the next section we will describe the toy model we adopt and estimate the typical values of the parameters which would fit the astronomical observations.  Then we will describe the proposed experimental set up required to make the observations.  Finally we will make some concluding remarks.

\section{Fluorescent Dark Matter}
Fluorescent Dark Matter describes a class of dark matter models that resonantly absorb and re-emit photons of a specific energy or energies - in our case, 3.5 keV. Its phenomenological features include the appearance of absorption and emission lines at the resonant energy that cannot be explained within the SM. As described above, this effect could be used to explain astrophysical anomalies at 3.5 keV. Anomalous scattering of photons at the resonant energy is also a generic feature of Fluorescent Dark Matter, and it is this effect that we search for in this paper.

A minimal Lagrangian for Fluorescent Dark Matter is:
\begin{equation}
\label{FDMlagrangian}
\mathcal{L} \supset \frac{1}{M} \bar{\chi}_{2} \sigma_{\mu \nu}  \chi_1 F^{\mu \nu} - m_1 \bar{\chi}_1 \chi_1 - m_2 \bar{\chi}_2 \chi_2,
\end{equation}
where $F_{\mu\nu}$ is the usual electromagnetic field strength and $\chi_1$ and $\chi_2$ are new fermions. We assume that $\chi_1$ makes up all the dark matter, while $\chi_2$ is a short-lived excited state. The process $\chi_1 \gamma \to \chi_2 \to \chi_1 \gamma$ occurs resonantly at photon energy $E_0 \stackrel{!}{=} \frac{m_2^2 - m_1^2}{2 m_1} = 3.54 \, {\rm keV}$ in the rest frame of the dark matter \cite{160801684}. Near resonance, the $\chi_1 \gamma$ interaction cross section is given by the relativistic Breit-Wigner formula. Assuming $\chi_2$ decays only to $\chi_1 \gamma$, we have a branching ratio of unity and the cross section reads \cite{Renard:1982bb}
\begin{equation}
\sigma_{\rm BW} (s) = \frac{2 \pi} {p_{\rm CM}^2} \frac{(m_2 \Gamma)^2}{(s - m_2^2)^2 + (m_2 \Gamma)^2}~,
\label{BW}
\end{equation}
where $\Gamma$ is the decay rate of $\chi_2$; $p_{\rm CM}^2 = \frac{m_1^2 E^2}{m_1^2 + 2 m_1 E}$ is the magnitude of each particle's momentum in the centre of mass frame, and $\sqrt{s}$ is the centre of mass energy. We emphasize that the Breit-Wigner cross section, equation \eqref{BW}, could arise from a wide range of FDM Lagrangians, of which equation \eqref{FDMlagrangian} is one simple example. Our proposed experiment is sensitive to the Breit-Wigner interaction with 3.5 keV photons, irrespective of the underlying particle physics model at the level of the Lagrangian. We are therefore sensitive to all Fluorescent Dark Matter models that explain the 3.5 keV astrophysical anomalies.

Assuming the dark matter is non-relativistic, we have:
\begin{equation}
s\simeq m_1^2+2m_1E_\gamma(1-\beta_{\rm DM}\cos\theta),
\label{com}
\end{equation} 
where $\theta$ is the angle between the dark matter and photon trajectories and $\beta_{dm}=v_{\rm DM}/c$ is the velocity of the dark matter particle. Typically, $\beta_{\rm DM}\simeq 10^{-3}$ in objects such as Milky Way size galaxies and galaxy clusters, the dark matter velocities being spread out in something quite similar to a Maxwell Boltzmann distribution with width $\sigma_{\rm DM}^2\simeq 10^{-6} c^2$ \cite{Fairbairn:2012zs}.

Very often this small velocity $\beta_{\rm DM}$ is negligible when considering the scattering of relativistic particles off dark matter since it results in a very small change in centre of mass energy for a given value of $m_1$ and $E_\gamma$.  However in the situation of Fluorescent Dark Matter, this velocity spread is critical.  Inspection of equation (\ref{BW}) shows that a narrow width of the resonance $\Gamma$ is required to obtain a large cross section, however there are simply not enough photons at precisely the right energy required to exploit this enhanced cross section without the effective broadening of the resonance due to the distribution of different values of $\beta_{dm}$ in equation $\eqref{com}$. \\

In \cite{160801684} it is pointed out that the equivalent width of the dip in the central AGN is approximately equal to the dark matter velocity broadening width. (The equivalent width is the energy range of the background model required to obtain the number of photons that are missing in the dip.) This suggests that photons within an energy range set by the dark matter velocity broadening ($\sim$15 eV) are completely absorbed. Therefore, we are only able to place a lower bound on the intrinsic width of the resonant photon absorption \cite{160801684}:

\begin{equation}
\Gamma \gtrsim \left( \frac{m_1}{{\rm GeV}} \right) \times (1 - 10) \times 10^{-10} \, {\rm keV},
\label{Gamma}
\end{equation}

where the range accounts for uncertainties in the dark matter column density. We discuss possible constraints on the FDM parameter space and more specific models in the appendix.

\section{Laser Experiment to excite Fluorescent Dark Matter}
We wish to observe the resonant scattering of 3.5 keV photons with photons from an X-ray laser, such as LCLS-II, and the dark matter wind on Earth. Rather than detecting a dip in the spectrum, we will observe the anomalous photon scattering by detecting the scattered photons themselves.  In this section we will explain the nature of the proposed experiment.
\subsection{Outline of the Experiment}
 The effective interaction cross section between the dark matter and the photon is strongly influenced by both the energy distribution of the laser, and the broadening of the interaction cross section by the dark matter's velocity. The Tender X-ray Imaging instrument (TXI) at LCLS-II, which covers our required energy, has a FWHM bandwidth of $3 \times 10^{-3}$ which gives rise to a width of $\Delta E_{\rm laser}\sim 10$ eV at 3.54 keV \cite{TXI}. The interaction cross section is broadened by the dark matter velocity dispersion, and by the fact that the Earth's rotation will change the angle between the laser and the dark matter wind. We assume that the Solar System moves through the dark matter halo with a velocity $\beta_{DM} \sim 10^{-3}$. The local dark matter velocity dispersion is $\sigma_{DM} \sim 10^{-3} c$, giving line broadening of $\sim 3$ eV. Furthermore, the exact position of the resonance will depend on the orientation of the laser with respect to the dark matter wind i.e. on the orientation of the laser with respect to the velocity vector of the Earth relative to the Galaxy which changes throughout the day, and based on the time of year that the experiment is performed. The resonant energy will therefore change on a daily and an annual basis. We will now estimate the magnitude of this effect. Consider an extremal case in which the laser is in the same plane as the dark matter wind, so that the Earth's rotation takes it from parallel to anti-parallel with the wind over the course of a day. The change in the resonant photon energy between these two geometries is $\Delta E_{\rm res} = p_{\rm DM} \frac{m_2^2 - m_1^2}{m_1^2} = \frac{2 p_{\rm DM}}{m_1} E_0$, where $p_{\rm DM}$ is the momentum of the dark matter wind; $m_1$ and $m_2$ are the masses of the dark matter and excited states respectively, and $E_0$ is the resonant photon energy in the rest frame of the dark matter. This gives $\Delta E_{\rm res} \sim 7$ eV, assuming the dark matter is non-relativistic. As $\Delta E_{\rm res}$ is of the same order of magnitude as the local dark matter velocity broadening, we will not consider the shift in the resonant energy further in this work. We note in passing that the predicted daily and annual modulations in $E_{\rm res}$ could potentially be used to distinguish signal from noise. For the rest of this work, we will simply assume a total FWHM line broadening for the dark matter-photon interaction resonance of $\Delta E_{\rm line} = 5$ eV. In fact, as (fortuitously) $\Delta E_{\rm line} \sim \Delta E_{\rm laser}$, the exact value does not have a large impact on our results. 

We can now compute the average cross section for the interaction of a photon from the laser and a dark matter particle:

\begin{equation}
\sigma _{\rm av} = \int \sigma_{\rm broadened} (E) f(E) dE,
\end{equation}

where $ \sigma_{\rm broadened} (E) $ is the velocity broadened cross section and $f(E)$ is the spectral shape of the laser, normalised such that $\int f(E) dE = 1$. We assume that both are Gaussian:

\begin{equation}
\sigma_{\rm broadened} (E) = A \, {\rm e} ^{ - \frac{1}{2} \left( \frac{E -E_0}{\sigma_{\rm line}} \right)^2},
\end{equation} 

where $\sigma_{\rm line} = \frac{\Delta E_{\rm line}}{2 \sqrt{2 {\rm ln}(2)}}$ and $A$ is chosen such that \\
\noindent $\int \sigma_{\rm broadened} (E) dE = \int \sigma_{\rm BW} (E) dE$.

\begin{equation}
f(E) = \frac{1}{\sigma_{\rm laser} \sqrt{2 \pi}} {\rm e} ^{ - \frac{1}{2} \left( \frac{E -E_0 - \delta E}{\sigma_{\rm laser}} \right)^2},
\end{equation}
where $\sigma_{\rm laser} = \frac{\Delta E_{\rm laser}}{2 \sqrt{2 {\rm ln}(2)}}$ and $\delta E$ is the offset of the laser's central energy from the resonant energy of the dark matter-photon interaction in the dark matter rest frame.

The rate of photon scattering from the laser is then given by:

\begin{equation}
R =  \sigma_{\rm av} n_{{\rm DM}} L F
\end{equation}
where $n_{{\rm DM}}$ is the number density of dark matter particles; $L$ is the length of laser over which we can detect the scattered photons, and $F$ is the photon flux. We assume a local dark matter density of 0.4 GeV cm$^{-3}$. The average X-ray power at LCLS-II will be limited to 200 W \cite{LCLSIITech}. At 3.54 keV, this corresponds to $F = 3.6 \times 10^{17}$ photons s$^{-1}$, which we use throughout the rest of this paper. We take $\delta E = 3$ eV. The resonant energy is currently known only to an accuracy of $\pm 20$ eV, but future observations, if they confirm the presence of the line, will improve on this. As discussed below, we assume $L = 7$ cm and a minimum decay width $\Gamma = \left( \frac{m_1}{{\rm GeV}} \right) 1 \times 10^{-10} \, {\rm keV}$. We obtain $R  = 1$  photon per week. Note that $R$ is independent of the dark matter mass, as the scaling of the required decay width and the dark matter number density exactly cancel.\\

Having seen that the numbers are not completely insignificant, in the next section we will describe in more detail a potential experimental set up to detect these photons.

\subsection{Experimental Set-up}

As explained above, we intend to observe a region through which an X-ray laser passes to see if the dark matter which passes through the same region becomes excited and re-emits photons in a different direction to the beam direction.

\begin{figure*}
\includegraphics[scale=0.7]{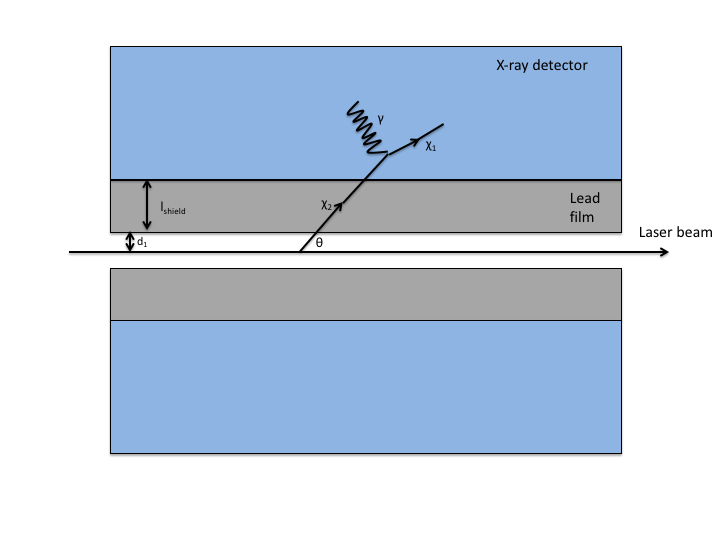}
\caption{Scattering of excited dark matter from a 3.5 keV laser, and subsequent decay to photons. (Not to scale.)}
\label{shielding}
\end{figure*}

There is a substantial background to this signal from Rayleigh scattering of the laser by air or other particles in the beam line. (Inelastic scattering of the laser is not problematic, as we can reject any photon detections with energies outside $E_{0} \pm \Delta E_{\rm line}$.) To overcome this background, we add a ``light shining through a wall'' component to our experiment. Between the laser beam and the X-ray detector, we place a material opaque to X-rays such as lead. X-ray photons scattered by any air remaining in the laser's path, or by any other effect, will therefore not reach our detectors. However, an on shell $\chi_2$ particle will pass through the lead shield before decaying to $\chi_1$ and a photon which may be observed in the detector. This is illustrated schematically in figure \ref{shielding}.

\begin{figure}
\includegraphics[scale=0.5]{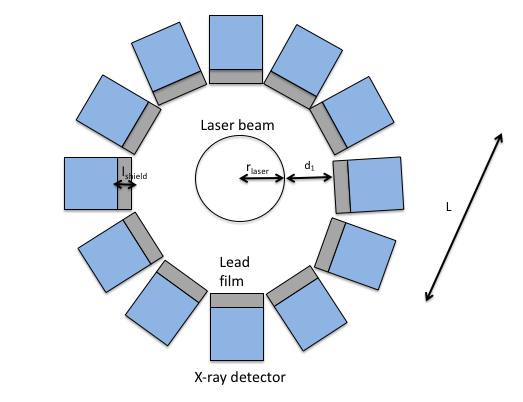}
\caption{A cross section of our experimental set-up (not to scale).}
\label{faceon}
\end{figure}

Our experimental set up is shown in figure \ref{faceon}. The laser beam is surrounded by a cylinder of X-ray detectors, each coated in a lead film to shield the detector from backgrounds. A suitable X-ray detector would be, for example, Medipix \cite{medipix}. The sides of the detector not adjacent to the laser beam would also need to be shielded with lead (or another material opaque to X-rays) to protect them from stray photons from the laser and other backgrounds. 

\begin{figure*}
\includegraphics[scale=0.7]{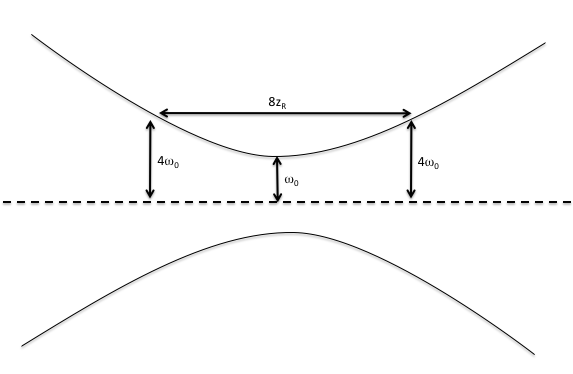}
\caption{The change in beam radius over 8 Rayleigh lengths. The beam waist, $\omega_0$ is the point with the smallest radius.}
\label{Rayleigh}
\end{figure*}

To determine the dimensions of the experimental apparatus, we must consider the Rayleigh length of the laser, $z_R$. $z_R$ is the length over which the cross sectional area of the beam doubles. We consider a length of beam over which the radius increases by a factor of 4 from its smallest value (the beam waist). This allows us to instrument over 4 Rayleigh lengths each side of the beam waist, as shown in figure \ref{Rayleigh}. For a beam waist of $\omega_0 \sim 1 \, \mu {\rm m}$ (within the capabilities of, for example, the TXI at LCLS-II), the Rayleigh length is:

\begin{equation}
z_{R} = \frac{\pi \omega_{0}^2}{\lambda} \simeq 9 \, {\rm mm}
\end{equation}

We therefore instrument a length $L = 8 z_{R} \simeq 7 \, {\rm cm}$. At its widest point in our apparatus, the beam's radius will be $r_{\rm laser} \sim 4 \, \mu {\rm m}$. To avoid damaging the shield with the tail of the laser's beam profile, we take a laser-shield distance $d_1 = 20 \, \mu {\rm m}$. The mass attenuation coefficient of 3.5 keV X-rays in lead is $\mu_{\rm lead} = 1.5 \times 10^3 \, {\rm cm}^2 {\rm g}^{-1}$ \cite{lead}, and its density at room temperature is $\rho_{\rm lead} = 11 \, {\rm g} \, {\rm cm}^{-3}$. We will take our lead film to have thickness $l_{\rm shield} = 3 \times 10^{-3}$ cm, such that it transmits only a fraction $\sim 10^{-22}$ of 3.5 keV photons incident directly on it. 

We now compute how our expected signal is modified by the detector's lead shield. For $\Gamma \simeq 10^{-10}$ keV, the lifetime of $\chi_2$ is $\tau \simeq 10^{-8}$ s. Given a dark matter velocity $v_{\rm DM} \simeq 10^{-3} c$, the particle travels an average distance $d \simeq 10^{-3}$ m before decaying - comfortably high enough to reach the detector. As $\Gamma$ increases, the $\gamma \chi_1$ interaction cross section increases, but the $\chi_2$ lifetime decreases, and so a smaller proportion of the excited dark matter scattered from the beam reaches the X-ray detector before decaying. Photons produced before $\chi_2$ reaches the detector will be blocked by the lead shield and therefore not contribute to our signal. The dependence of the expected signal on the resonant interaction width $\Gamma$ is therefore not monotonic. In particular, for higher mass dark matter, the required interaction cross section is greater, and so $\tau$ must decrease even when $\Gamma$ is at the minimum value required to explain the anomaly observed in NGC 1275. The proportion of $\chi_2$ particles produced by the beam that pass through the lead shield before decaying depends in the their angular distribution - particles produced at smaller angles to the laser beam will have a longer path through the shield. 
This distribution depends most significantly on the orientation of the laser with respect to the dark matter wind, which will change on a daily and on an annual basis. We therefore assume that, averaged over the running time of the experiment, the angular distribution is uniform.

The average rate of detected photons is given by: 

\begin{equation}
R = \int \frac{d \sigma_{\rm av}}{d \Omega} n_{\rm DM} L F {\rm e}^{-t (\Omega) / \tau} d \Omega
\end{equation}

Assuming a uniform differential cross section, we take $\frac{d \sigma_{\rm av}}{d \Omega} = \frac{\sigma_{\rm av}}{4 \pi}$. We will use spherical polar coordinates with the laser's direction as the z axis. We then have $t(\Omega) = \frac{d_1 + l_{\rm shield}}{v_{\rm DM} {\rm sin}(\theta)}$ is the time taken for $\chi_2$ to traverse the lead shield. $d_1 = 20 \, \mu $m is the distance between the laser and the lead shield, and $l_{\rm shield} = 30 \mu $m is the size of the lead shield. We neglect the finite size of the laser in this estimate. We therefore obtain:

\begin{equation}
R = \frac{1}{2}  \int_0^{\pi} d \theta \, {\rm sin}( \theta ) \sigma_{\rm av} n_{\rm DM} L F {\rm exp} \left( \frac{-(d_1 + l_{\rm shield})}{ v_{\rm DM} \tau {\rm sin}( \theta )} \right) 
\end{equation}

\begin{figure*}
\includegraphics[scale=0.7]{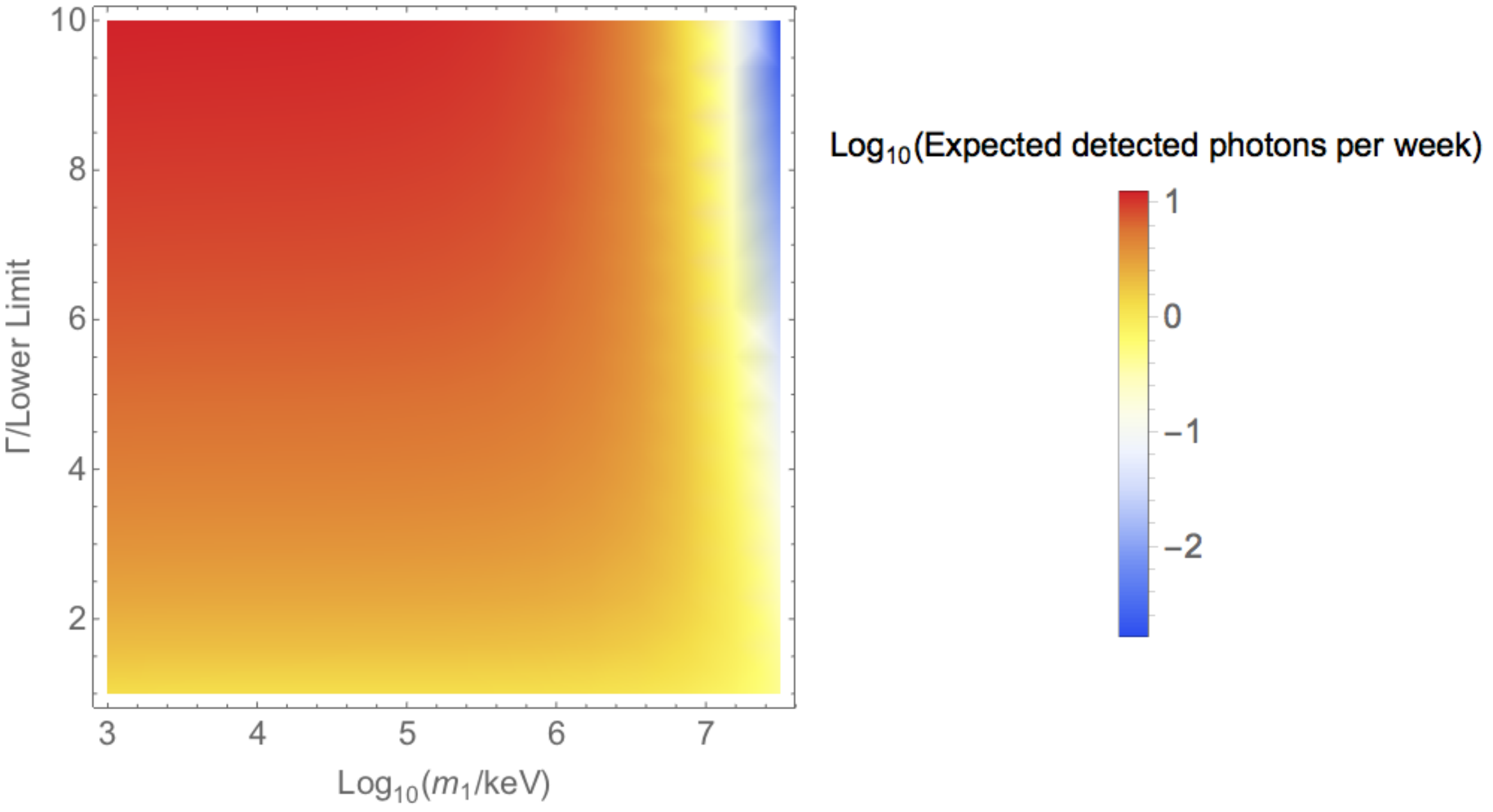}
\caption{The expected signal from a 3.5 keV laser scattering from Fluorescent Dark Matter. The vertical axis shows the ratio of the resonant width for the dark matter-photon interaction to the minimum value required to explain observed astrophysical anomalies (see equation \eqref{Gamma}).}
\label{signalPlot}
\end{figure*}

The expected rate of scattered photons reaching the detector depends non-linearly on the width $\Gamma$ of the resonant interaction through both $\sigma_{\rm av}$ and $\tau$. This dependence is shown in figure \ref{signalPlot}.

We must also consider the possibility that $\chi_2$ could pass through the detector completely before decaying, such that the resulting photon would not be measured. The largest $\chi_2$ lifetime shown in in figure \ref{signalPlot} corresponds to a distance of $\sim 2$ m for speed of $10^{-3} c$. Here we assume that several meters of detector may be built around the beam, either continuously or by stacking smaller detectors. If the detector of significantly smaller than this, the sensitivity in the lower mass range would be greatly reduced.

We now calculate the expected background from Rayleigh scattering. We will assume the experiment is carried out in an ultra-high vacuum of $10^{6}$ particles per ${\rm cm}^{3}$ of dry air, giving a density of $\rho_{\rm air} =  5 \times 10^{-17} \, {\rm g} \, {\rm cm}^{-3}$. The interaction coefficient for Rayleigh scattering of 3.54 keV photons from dry air is $\mu_{\rm air} = 0.75 \, {\rm cm}^{2} \, {\rm g}^{-1}$ \cite{lead}. The rate of photons from Rayleigh scattering reaching the detector is then:

\begin{equation}
\begin{split}
R_{\rm background} & \simeq \mu_{\rm air} L \rho_{\rm air} F {\rm e}^{- \mu_{\rm lead} l_{\rm shield} \rho_{\rm lead}}  \\
& = 10^{-15} \, {\rm photons} \, {\rm per} \, {\rm week}
\end{split}
\end{equation}

This background may therefore be safely neglected. The background rate could also be checked experimentally by running at an energy around 100 eV from the resonant energy, where the signal from resonant dark matter scattering is predicted to be negligible, but other backgrounds should be very similar.\\

A potentially observable signal is predicted for a wide range of Fluorescent Dark Matter parameters for an LCLS-II like laser. 






\section{Conclusion}

In this work we have presented a new way of searching for dark matter that can be excited by photons in the keV range. Such dark matter is motivated by recent astrophysical anomalies. If this Fluorescent Dark Matter exists, it will be excited by X-ray laser photons and subsequently decay, re-emitting a photon in a direction no longer along the beam line and from a displaced vertex.  We seek to search for dark matter of this nature by instrumenting a length of X-ray laser with detectors. 

We have presented an idealised experimental set up where photons are absorbed by dark matter along the beam line, the excited dark matter then passes through a lead shield surrounding the beam, before decaying on the other side, which is instrumented with detectors.  The lead shield is designed to block out X-rays scattered by normal molecules in the cavity.

Our calculations show that it might be possible to detect of order 1 - 10 photons a week in this way, thereby potentially opening up a new search strategy for dark matter.

Finally we point out that a similar strategy could in principle be used to search for FDM with a resonant energy anywhere on the electromagnetic spectrum. 

\section*{Acknowledgements}
The work of FD has been partially supported by STFC consolidated grant ST/P000681/1, and by Peterhouse, University of Cambridge. The work of MF was supported partly by the STFC Grant ST/L000326/1 and also by the European Research Council under the European Union's Horizon 2020 program (ERC Grant Agreement no. 648680 DARKHORIZONS). We are indebted to Rob Shalloo for sharing his expertise on lasers. We also thank Joe Conlon, Sven Krippendorf, Axel Linder, Michael Peskin and Andreas Ringwald for invaluable discussions. 

\section*{Appendix:- Constraints and Potential Models of Fluorescent Dark Matter}

Here we discuss in slightly more detail the kind of models which would lead to a signal using the methods in this paper.

In \cite{160304859} a more general version of equation \eqref{FDMlagrangian}, including both electric and magnetic dipole interactions, is used to explain the 3.5 keV line. For the electric dipole interaction $\mathcal{L} \supset \frac{1}{M'} \bar{\chi}_{2} \sigma_{\mu \nu} i \gamma^5 \chi_1 F^{\mu \nu}$, \cite{160304859} finds that the dominant contribution to the 3.5 keV line arises from upscattering of $\chi_1$ from the intracluster plasma, rather than from resonant photon absorption. This process contributes to the line signal but not to the dip signal, and would not lead to a signal in the experiment described in this work. We therefore restrict ourselves to considering the effects of resonant dark matter-photon scattering, with the Breit-Wigner cross section as our starting point.

It is also shown in \cite{160304859} that consistency with LEP constraints on the running of $\alpha$ requires $M \gtrsim 200$ GeV. Conversely, explaining the 3.5 keV dip with equation \eqref{FDMlagrangian} requires $M \lesssim 10$ GeV, given $\Gamma \simeq \frac{(m_1 - m_2)^3}{\pi M^2}$ and taking $m_1 \gtrsim 1$ MeV as required for the consistency of the dip and line energies \cite{160304859,160801684}. This rules out a dark sector with two fermions coupled to the photon by a dimension five dipole operator, with a coupling strong enough to explain the dip and with no other relevant new physics coming into play. However, if the hidden sector is more complicated than this, the bound could be different when the full UV theory is considered. For example, there could be other new physics that cancels the contribution from the dipole operator. 


One possibility for such a theory with a more complicated hidden sector is millicharged DM with a hidden strongly coupled gauge group. Indeed, LEP data constrains loop corrections to $\alpha$ at a level of $\sim 20 \%$ \cite{hep-ex/0507078}, and therefore relatively high millicharges are still allowed by this particular constraint. We might therefore seek a scenario where bound states of new particles with allowed millicharges give the decay rate required to explain the dip. Both the bounds that must be evaded and the transition rates between dark composite states will depend on the details of the model, such as the strong gauge group and the number of new fermions. For example, bounds on millicharged particles might be more easily evaded if the lightest composite state is neutral, and if the dark hidden sector is confining. We emphasise that the purpose of this paper is to see whether the signal apparently observed in Perseus, which is quite naturally interpreted as anomalous scattering of 3.5 keV photons, is also observed in an Earth based experiment. This question is completely insensitive to the UV physics.

As a simple example, we might consider a theory in which the dark sector consists of at least one new fermion $f$, charged under a hidden gauge group $U(1)'$ and a hidden strongly interacting non-Abelian gauge group. The dark photon corresponding to $U(1)'$ is assumed to be massive, with a small kinetic mixing $\epsilon$ with the Standard Model photon. A sample Lagrangian for this dark sector is:

\begin{equation}
\label{darkLagrangian}
\begin{split}
\mathcal{L} \supset & \bar{f} (i \slashed D'_{\mu} - m_f)f - \frac{1}{4} G'^a_{\mu \nu} G'^{\mu \nu}_a - \frac{1}{4} F'_{\mu \nu} F'^{\mu \nu} \\ & - \frac{\epsilon}{2} F'_{\mu \nu} F^{\mu \nu} - \frac{1}{2} m'^2 A'_{\mu} A'^{\mu}, 
\end{split}
\end{equation}

where $G'$ is the dark gluon field strength tensor, $A'$ is the dark photon with $F'$ its field strength tensor and $D'$ is the covariant derivative with respect to both hidden gauge groups. As with strong interactions in the SM, we expect the formation of composite dark states. We assume the lowest energy composite state is electrically neutral with mass $m_1$ and comprises the dark matter, with an excited state with mass $m_2$. We take $E_0 \stackrel{!}{=} \frac{m_2^2 - m_1^2}{2 m_1} = 3.54 \, {\rm keV}$, as above.

The dynamics of the strong dark sector depend on the dark confinement scale $\Lambda'$. In the case $m_f \ll \Lambda'$, $m_1$ is set by $\Lambda'$. Conversely, when $m_f \gg \Lambda'$, $m_1$ is set by $m_f$. In either case, we assume $m' > m_2$, so that decay of $m_2$ to $m_1$ and a dark photon is kinematically forbidden. The excited dark matter therefore decays with $100 \%$ branching ratio to $m_1$ and a Standard Model photon, through the kinetic mixing between the dark and SM photon.  We could therefore reproduce the interaction described by the Breit-Wigner resonance in equation \eqref{BW}. 

Calculating the predictions of and bounds on the class of models described above would be a significant undertaking. Given the compelling astrophysical evidence for the anomalous resonant scattering of 3.5 keV photons, we will go on to describe an Earth based experiment to detect this same resonant scattering, and defer detailed discussion of possible UV completions that may evade the bounds given in \cite{160304859} to later work. The experiment proposed in this work tests the very same interaction - resonant scattering between dark matter and 3.5 keV photons - as that proposed to explain the 3.5 keV line and dip, and is therefore insensitive to the UV theory generating these interactions.\\

\bibliographystyle{unsrt}
\bibliography{laser}

\end{document}